\documentclass[prl,twocolumn,aps,floats,showpacs]{revtex4}
\usepackage{graphicx}
\usepackage{psfrag}

\input{tcilatex}

\begin{document}

\title{Quantum nonlocality obtained from local states by entanglement
purification}
\author{P. Walther$^{1\ast }$, K.J. Resch$^{1\ast }$, \v{C}. Brukner$^{1}$,
A.M. Steinberg$^{1,2}$, J.-W. Pan$^{1,3}$, and A. Zeilinger$^{1,4}$}
\affiliation{$^{1}$Institut f\"{u}r Experimentalphysik, Universit\"{a}t Wien,
Boltzmanngasse 5, 1090 Vienna, Austria\\
$^{2}$Department of Physics, University of Toronto, 60 St. George Street,
Toronto ON\ M5S 1A7 Canada\\
$^{3}$Physikalisches Institut, Universit\"{a}t Heidelberg, D-69120
Heidelberg, Germany\\
$^{4}$IQOQI,Institut f\"{u}r Quantenoptik und Quanteninformation, \"{O}%
sterreichische Akademie der Wissenschaften, Austria\\
$^{\ast }$ These authors contributed equally to this work.}

\begin{abstract}
\textbf{We have applied an entanglement purification protocol to
produce a single entangled pair of photons capable of violating a
CHSH Bell inequality from two pairs that individually could not.
The initial poorly-entangled photons were created by a
controllable decoherence that introduced complex errors.  All of
the states were reconstructed using quantum state tomography
which allowed for a quantitative description of the improvement
of the state after purification.}
\end{abstract}

 \maketitle

Distributed entanglement is a prerequisite for quantum
communication and quantum computation. Unfortunately,
entanglement is very fragile and can prove difficult to maintain
particularly between separated parties and over extended
timescales. This places stringent limits on the range over which
quantum communication can take place without some sort of
restoration. In this experimental work, we create two photon
pairs with low entanglement and purity through controllable
decoherence. This noise is very complex and is not engineered
with a specific, but limited, purification scheme in
mind~\cite{kwiatconcent,yamamoto}. After this decoherence each
pair cannot violate a CHSH inequality~\cite{CHSHBell1} or any
other known inequality \cite{collins04} and we refer to the
states as ``local''. We apply an entanglement purification
protocol to these initial pairs and through quantum state
tomography show quantitative improvement in the qualities of the
purified pair. This is confirmed by the final pair violating a
CHSH Bell inequality proving that it cannot be described by a
local realistic model and can serve as the crucial entanglement
resource for a number of quantum communication protocols, such as
quantum communication complexity~\cite{collins02}.

Entanglement can be created via local interactions between particles.
However, once the particles are no longer in direct contact, or connected by
a quantum channel, their entanglement cannot be increased on average~\cite%
{loccrules}. There are, however, methods for increasing the entanglement or
state purity of a subset of the entangled particles~\cite%
{hiddennonlocality1,hiddennonlocality2}. In the general entanglement
purification scheme proposed by Bennett et al.~\cite{bennettpurification},
Alice and Bob share two pairs of low-quality entangled particles such that
each has one particle from each pair. They perform local entangling
operations on their respective particles, and then each make a local
measurement on one of the particles. By comparing their measurement outcomes
and performing local transformations based on those results, they retain a
single shared pair of particles which can be of both higher entanglement and
higher purity. After many rounds of this procedure, Alice and Bob can share
a state asymptotically close to a maximally-entangled pure state such as one
of the Bell states $|\phi^{\pm}\rangle=\frac{1}{\sqrt{2}}(|H\rangle|H\rangle%
\pm|V\rangle|V\rangle)$ and $|\psi^{\pm}\rangle=\frac{1}{\sqrt{2}}%
(|H\rangle|V\rangle\pm|V\rangle|H\rangle)$. While this scheme requires
technically difficult CNOT operations~\cite{cnotexp1}, it was shown that a
commonly used linear optical element, the polarizing beam-splitter (PBS),
could be used in its place if one is satisfied with probabilistic operation~%
\cite{jianweipurtheory}.

The entanglement purification experiment of Pan et
al.~\cite{jianweipurexp} served as a proof of principle, but the
qualities of the input states were reduced in an artificial way.
Input states were created via local polarization rotations on
highly pure and entangled photon pairs. By averaging over two
different polarization rotations, the authors argued they had an
effectively mixed state. However, local polarization rotations
cannot, change either the purity or entanglement of the photon
pairs. Furthermore, these rotations only introduce bit-flip
errors and create mixtures of only two of the four Bell states.
Mixtures of two Bell states always display nonlocal behaviour
(i.e., violate a Bell inequality), except for the case of exactly
equal mixture~\cite{horodecki}, which cannot be purified.

Here, we create effectively mixed polarization states by
entangling the polarization degree of freedom from highly
entangled photon pairs with their arrival times~\cite%
{kwiatconcent,hiddennonlocality2,dfstomog}; these time shifts are
far too small to be measured by our detectors and constitute for
all practical purposes irreversible interaction with the
environment. Entanglement with this unobserved degree of freedom
reduces the quantum coherences in the polarization states
required for high entanglement and purity. Our decoherence
creates states with incoherent contributions from all of the Bell
states by introducing complex phase, bit-flip, and correlated
errors. Using quantum state tomography~\cite{tomography}, we
reconstruct the full two-photon polarization states of our
initial pairs and purified pair. The Clauser-Holt-Shimony-Horne
(CHSH) Bell inequality~\cite{CHSHBell1} shows a conflict between
local realism and measured polarization correlations when the Bell
parameter, $S$ (a function of those correlations), is larger than
2. We introduce sufficient decoherence to make our initial states
incapable of violating such an inequality for any analyzer
settings. We then quantitatively measure the improvement in the
entanglement and the purity of our purified state; our final
state violates a Bell inequality by more than 2 $\sigma$.

At the heart of this entanglement purification method is the PBS
(Fig. 1). A PBS transmits horizontally-polarized (H) light and
reflects vertically-polarized (V) light. If two photons enter the
PBS from two different inputs, then the photons are sent to
different outputs if and only if they are both H or both V. For
this reason the action of the PBS and post-selection of photons
in the two different outputs constitutes a determination that the
parity of the photons is even~\cite{jianweipurtheory,parity}. The
CNOT operation on a control (C) and target (T) photon
polarization performs: $|H\rangle_C|H\rangle_T\rightarrow|H\rangle_C|H\rangle_T$, $%
|H\rangle_C|V\rangle_T\rightarrow|H\rangle_C|V\rangle_T$, $%
|V\rangle_C|H\rangle_T\rightarrow|V\rangle_C|V\rangle_T$, $%
|V\rangle_C|V\rangle_T\rightarrow|V\rangle_C|H\rangle_T$. These
operations are related since if the target photon is measured in
the state H after a CNOT, then the input photons had even parity.
However, the PBS-based scheme is limited to $50\%$ of the
efficiency of the CNOT scheme, since it only measures cases with
even parity. Furthermore, the linear optics parity check only
generates the requisite entanglement when the photons overlap
coherently at the PBS.

In our protocol (Fig. 1) Alice and Bob each has one photon from
each entangled pair. They, locally, perform parity checks and
measure the polarization of one of their PBS outcomes (modes a1'
\& b2') in the $|\pm \rangle =1/\sqrt{2}(|H\rangle \pm
|V\rangle)$ basis. Alice and Bob keep the photons in the other
modes (a2' \& b1'), only in those cases when they both obtain the
$|+\rangle$ outcome. If each pair of entangled photons has a
minimum fidelity $F>0.5$ with a Bell state, and if both parity
checks succeed, then those remaining photons can be both of
higher purity and of higher entanglement~\cite{jianweipurtheory}.
The interested reader is referred to~\cite{bennettpurification}
for a mathematical analysis of purification using CNOT operations
and to~\cite{jianweipurtheory,jianweisimontheory} for a detailed
theoretical treatment of the linear optics implementation.

In our experiment, we create two pairs of entangled photons using
type-II parametric down-conversion~\cite{kwiatsource}. An
ultraviolet (UV) pulse passes twice through a $\beta$-barium
borate (BBO) crystal, which emits highly-entangled photons both
into the forward pair of modes a2 \& b2 and into the backward
pair of modes a1 \& b1 (Fig. 2). To counter the effects of
birefringence in the down-conversion BBO crystal, one normally
rotates the polarization of each photon by $90^{\circ }$ with
half-wave plates (HWP) and passes them through a second set of BBO
crystals. Failure to do so results in degradation of the pair's
entanglement from unwanted correlations between the polarization
and the time-of-arrival of the photons. By rotating by an angle
less than $90^{\circ }$, we introduce controllable effective
decoherence in the polarization states of our photon pairs.

Down-conversion is a probabilistic source of entangled photons
and thus can produce two photon pairs in the same pair of modes
(eg. a2 \& b2) with about the same probability as two pairs into
four different modes. It has been shown that if the phases of the
forward double pair and backward double pair are held stable,
then the double pair emission can enhance the efficiency of
purification over the case with no double pair
emission~\cite{jianweisimontheory}. We, therefore, kept the path
lengths strictly controlled and enclosed the setup to minimize air
fluctuations for the duration of the measurements.

The density matrices of the forward- and backward-emitted pairs
were reconstructed via a ``maximum likelihood'' method
\cite{maxlike,maxlikelihood} from our experimental data are shown
in Fig. 3a and 3b, respectively. The HWP in each decoherer
rotated the photon polarizations by $50^{\circ }$ for the forward pair and $%
62^{\circ }$ for the backward pair, instead of the ideal $90^{\circ }$.
These rotation angles were chosen to reduce the maximum possible Bell
parameter to $S<2$. The angles differ most likely due to experimental
asymmetries in the pump characteristics and to spatial filtering. The two
initial density matrices are similar, but have interesting differences. Both
input pairs contain large diagonal elements in the HH and VV positions with
non-maximal negative coherences so both states are primarily $|\phi
^{-}\rangle $. The extra diagonal terms and coherences in the density
matrices, more prominent for the forward pair, indicate that there are also $%
|\psi ^{\pm }\rangle $ components in our states. The smaller than
maximal coherences between HH and VV, which are more prominent
for the backward pair, show that our states are highly mixed.
Using the method of Horodecki et al.~\cite{horodecki} on our
initial states, we find that the maximum CHSH-Bell parameter one
could possibly measure is $S_{MAX}=1.89\pm 0.02$ for the forward
pair and $S_{MAX}=1.90\pm 0.014$ for the backward pair. The
errors on quantities extracted from the density matrices were
calculated via a Monte Carlo procedure. Both of our initial
states are more than 5 $\sigma $ below the nonlocality border of
$S=2$  and cannot serve as the crucial entanglement resource for
many quantum communication protocols. Our initial states were
also checked against the 3322 and 3422 inequalities of Collins
and Gisin~\cite{collins04}, which are complementary to the CHSH
inequality, and found that these inequalities could also not be
violated.

The PBS acts in a preferred basis, the H/V basis. This makes purification
more efficient at purifying Bell states with different H/V correlations
(i.e., $|\psi ^{\pm }\rangle $ states from $|\phi ^{\pm }\rangle $ noise)
than it is from purifying Bell states with the same H/V correlations (i.e.,$%
|\psi ^{\pm }\rangle $ states from $|\psi ^{\pm }\rangle $
noise). One can convert between different Bell states through
local polarization rotations without changing the entanglement or
purity. With coloured noise, these local rotations can increase
the efficiency of purification. We perform polarization rotations
using the four HWP which convert $|\phi^{-}\rangle $ to
$|\psi^{+}\rangle$ and also permute the noise contributions. This
allows higher degree of purification and converts our output
state to $|\psi^{+}\rangle $.

Fig. 3c shows the density matrix of our final state after the
purification protocol. Whereas the initial state measurements
were based on direct 2-photon coincidence counts, the final state
measurements were based on 4-fold coincidence counts - two counts
signal that Alice and Bob both measure the $|+\rangle $ outcome
and the second two are used for tomography of the purified state.
Four-fold coincidences were accumulated for 6000s per point and
were normalized by the square of the coincidences between the two
fixed measurement detectors (modes a1' \& b2') to account for
changes in the laser power. The final state contains large
diagonal elements in the HV and VH positions and has large
positive coherences between them - this is the signature of
$|\psi ^{+}\rangle $. The larger coherences and the lack of
significant terms in other diagonal density matrix elements
indicate qualitatively that our state has both improved purity
and entanglement characteristics. We express the improvement
quantitatively by computing the maximum Bell parameter one could
measure from this state for optimal measurement settings,
$S_{MAX}=2.28\pm 0.06$. From our tomographic reconstruction we
conclude that our final state can violate a Bell inequality by
$4\sigma $. Using this same final state, we explicitly measured
the Bell parameter for the measurement settings of $-22.5^{\circ
}$, $22.5^{\circ }$ for Alice (mode a2') and $0^{\circ }$,
$45^{\circ }$ for Bob (mode b1') to be $S_{MEAS}=2.29\pm 0.13$ --
a $2.2\sigma $ violation. For the same settings, we
predict the Bell parameter from the density matrix and find $%
S_{DM}=2.17\pm 0.07$, in good agreement with our measured results.

For bipartite states, the entanglement can be quantified by the
tangle~\cite{tangle}, while the purity can be quantified by the
linear entropy $S_{L}=\frac{4}{3}(1-Tr[\rho^{2}])$. To show that
the entanglement and purity of our state are improved through the
purification process we plot the input and final states in Fig. 4
~\cite{tangleentplaneMEMS}. The initial states have very low
Tangle (low entanglement) and high Linear Entropy (low purity).
Both the tangle and entropy of our final state are significantly
improved by the purification process.

Decoherence is one of the most difficult obstacles facing
experimental quantum information processing and communication. In
this work, two photon pairs were decohered by complex noise that
introduced both phase and bit-flip errors. We have used a
linear-optics-based protocol to purify the states, and
quantitatively measured the resulting improvement in their
quantum properties. Our initial states were so noisy that their
polarization correlations could not violate a CHSH inequality and
would thus be useless for many quantum communication tasks. In
contrast, the nonlocality of our final state was explicitly
demonstrated by violation of a Bell inequality. The newly created
nonlocal correlations constituted an entanglement resource for
quantum communication protocols. Counteracting the negative
effects of decoherence is essential to the success of quantum
information.

The authors thank Chris Ellenor and Morgan Mitchell for expertise
and software and Andrew White for helpful discussions. This work
was supported by the Austrian Science Foundation (FWF), NSERC,
the DARPA\ QuIST\ program managed by the Air Force Office of
Scientific Research, and the European Commission (RAMBOQ).

\noindent \textbf{Fig. 1} Schematic for entanglement purification.
A source produces two pairs of entangled photons, one photon from
each pair travels to Alice and the others to Bob. In each quantum
channel, decoherence degrades the entanglement and purity of the
quantum states. Each party mixes their photons through a
polarizing beam-splitter (PBS) oriented in the H/V
basis and measures the outputs a1' \& b2' in the $|\pm \rangle =1/\sqrt{2}%
(|H\rangle \pm |V\rangle $ basis. Given that the four photons each take a
different PBS output and both Alice and Bob find the same measurement
outcome (in our experiment $|+\rangle $), the remaining photon pair in modes
a2' \& b1' has been purified.

\noindent \textbf{Fig. 2} The experimental setup for purifying
mixed entangled states. Entangled photon pairs are created when
an ultraviolet laser pulse makes two passes through a $\beta
$-barium borate (BBO) crystal. Rotating the half-wave plate (HWP)
in front of each compensating BBO crystal (COMP) by less than
$90^{\circ }$ creates a tunable degree of decoherence. All four
HWPs are left in to rotate $45^{\circ }$ to enhance the
efficiency of purification under our experimental noise
conditions. Modes from the two different pairs (a1 \& a2 and b1
\& b2) are combined at the two PBSs which perform parity checks.
In the output modes a1' \& b2', projective measurements are made
onto the state $|+\rangle $ using single-mode fiber coupled
single-photon counting detectors behind 3-nm bandwidth
interference filters. Tomographic and Bell inequality
measurements are performed using the HWPs and quarter-wave plates
(QWPs) in these output modes and two more fiber-coupled detectors
also behind interference filters.

\noindent \textbf{Fig. 3} Tomographic reconstruction of the
density matrices before and after purification. The real
(left-hand side) and imaginary (right-hand side) parts of the
density matrices of a) the initial forward pair, b) the initial
backward pair, and c) the purified pair. The maximum possible Bell parameter, $S_{MAX}$%
, from the CHSH-Bell inequality for any set of measurements is
$1.89\pm 0.02$ for the initial forward pair and $1.90\pm 0.014$
for the initial backward pair. For the purified state we measured
the Bell parameter with the polarizer settings $-22.5^{\circ }$ \&
$22.5^{\circ }$ in mode a3 and
$0^{\circ }$ \& $45^{\circ }$ in mode b4 to be $S_{MEAS}=2.29\pm 0.13>2$ -- a $2.2$-$%
\sigma $ Bell inequality violation. The final purified state
cannot be described by any local realistic theory.

\noindent \textbf{Fig. 4} Quantifing the state's improvement
through purification. Quantum state purity and entanglement can
be characterized by the linear entropy of the quantum state and
by its tangle. The initial states (open circle and open square)
are of high linear entropy and low tangle - i.e., they are highly
mixed and have low entanglement. After purification, the new
state (solid square) is of higher purity and higher entanglement.
For reference, all physical states lie below the solid
line~\cite{tangleentplaneMEMS}.

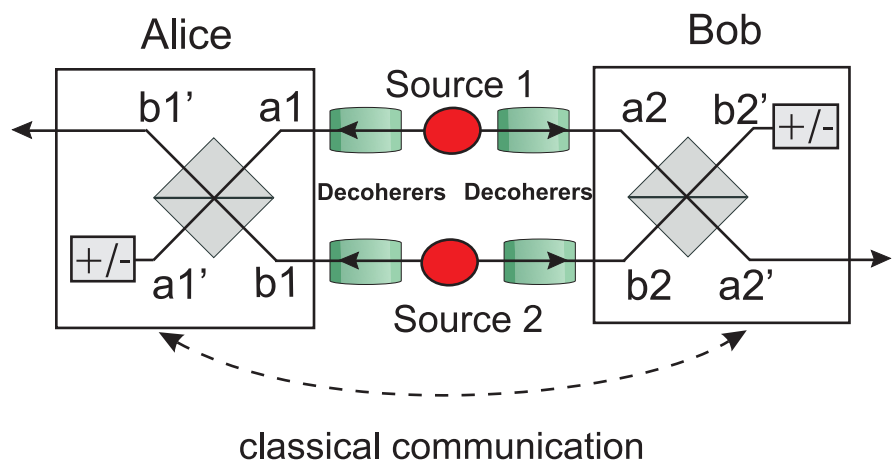
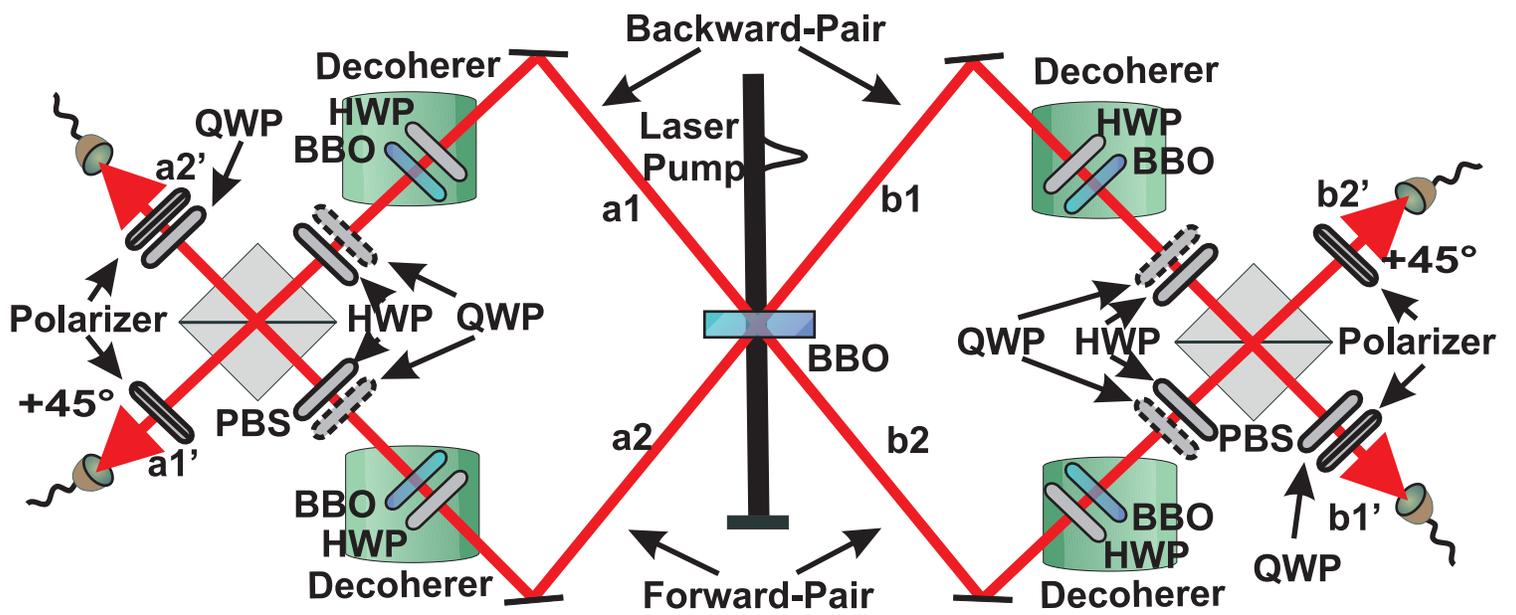
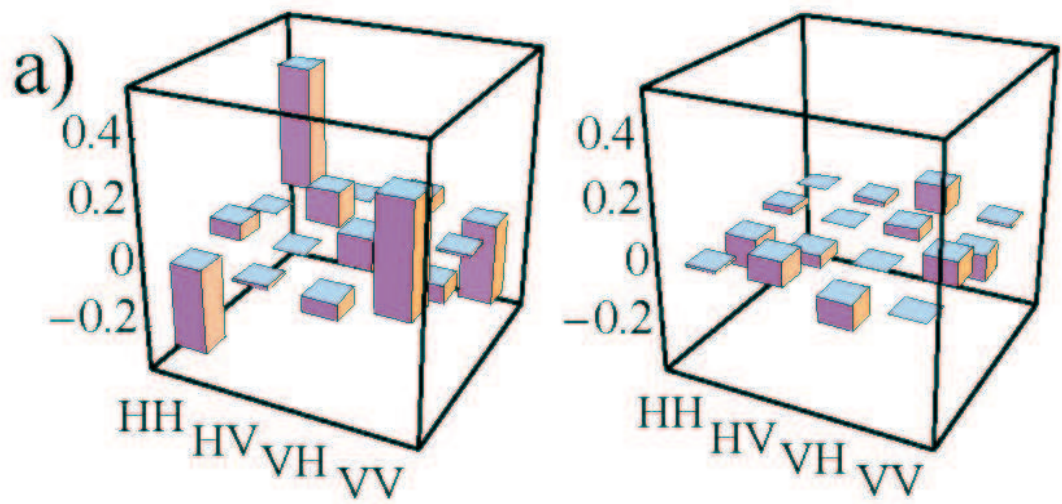
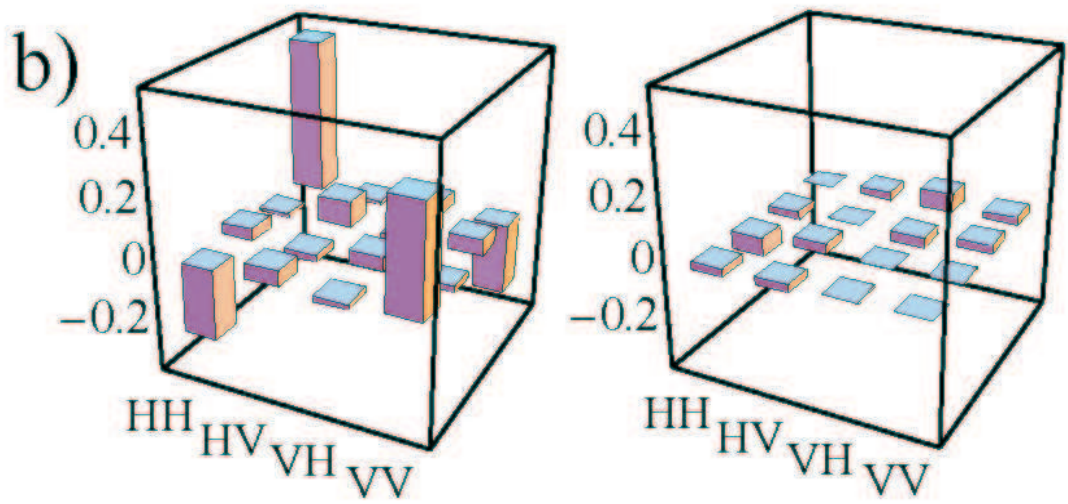
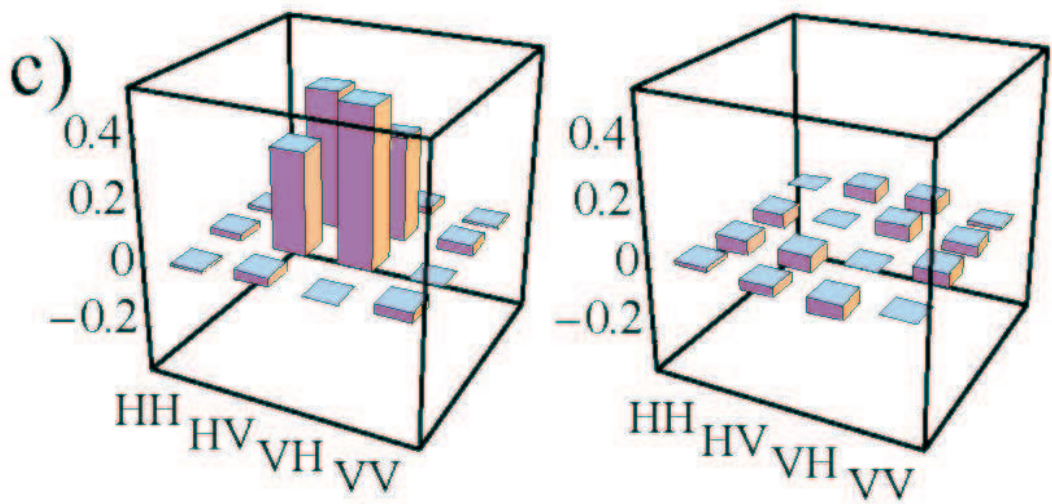
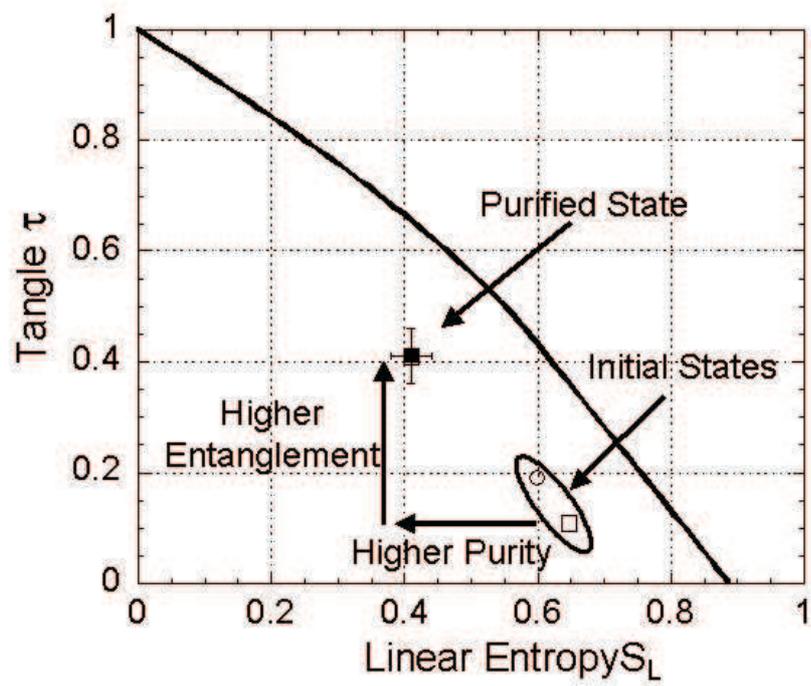
\end{document}